\documentclass[11pt,a4paper]{article}
\usepackage[T1]{fontenc}
\usepackage{amssymb}
\usepackage{mathptmx}
\usepackage[pdftex]{graphicx}
\usepackage[english]{babel}
\usepackage[pdftex,linkcolor=black,pdfborder={0 0 0}]{hyperref}
\usepackage{enumitem} % Includes lists
\usepackage{setspace}
\usepackage[a4paper, lmargin=0.12\paperwidth, rmargin=0.12\paperwidth, tmargin=0.1111\paperheight, bmargin=0.1111\paperheight]{geometry} %margins
\usepackage{wallpaper}
\usepackage{mdframed}
\usepackage{tikz}
\usepackage{comment}
\usepackage[affil-it]{authblk}
\usepackage[all]{nowidow} % Tries to remove widows
\usepackage{csquotes}
\usepackage{amsmath}

\author[1]{Nicola Melchioni}
\author[2, 4]{Federico Paolucci}
\author[3]{Paolo Marconcini}
\author[3]{Massimo Macucci}
\author[4]{Stefano Roddaro}
\author[4]{Alessandro Tredicucci}
\author[4,5,*]{Federica Bianco}

\affil[1]{NEST Laboratory, Scuola Normale Superiore, Piazza S. Silvestro 12, I-56127 Pisa, Italy}

\affil[2]{INFN Sezione di Pisa, Largo Bruno Pontecorvo 3, I-56127 Pisa, Italy}

\affil[3]{Dipartimento di Ingegneria dell’Informazione dell’Università di Pisa, via G. Caruso 16, I-56122 Pisa, Italy}

\affil[4]{Dipartimento di Fisica ``E. Fermi", Università di Pisa, Largo Pontecorvo 3, I-56127 Pisa, Italy}

\affil[5]{NEST Laboratory, Istituto Nanoscienze-CNR, Piazza S. Silvestro 12, I-56127 Pisa, Italy}

\affil[*]{Corresponding author: E-mail: federica.bianco@nano.cnr.it}

\begin{document}

\title{Tailoring coherent charge transport in graphene by deterministic defect generation}
\date{} %%%date is omitted
\maketitle

% Keywords: Please provide a minimum of three and a maximum of seven keywords, separated by commas

\begin{abstract}
Harnessing the wave-nature of charge carriers in solid state devices, electron optics investigates and exploits coherent phenomena, in analogy with optics and photonics. Typically, this requires complex electronic devices leveraging macroscopically coherent charge transport in two-dimensional electron gases and superconductors.
Here, collective coherent effects are induced in a simple counterintuitive architecture by defect engineering. Deterministically introduced lattice defects in graphene enable the phase coherent charge transport by playing the role of potential barriers, instead of scattering centres as conventionally considered. Thus, graphene preserves its quasi-ballistic quantum transport and can support phase-matched charge carrier-waves. Based on this approach, multiple electronic Fabry–P\'erot cavities are formed by creating periodically alternating defective and pristine nano-stripes through low-energy electron-beam irradiation. Indeed, defective stripes behave as partially reflecting mirrors and resonantly confine the charge carrier-waves within the pristine areas, giving rise to Fabry–P\'erot resonant modes. These modes experimentally manifest as sheet resistance oscillations, as also confirmed by Landauer-Buttiker simulations. Moreover, these coherent phenomena survive up to 30 K for both polarities of charge carriers, contrarily to traditional monopolar electrostatically created Fabry-P\'erot interferometers.
Our study positions defective graphene as an innovative platform for coherent electronic devices, with potential applications in nano and quantum technologies.
\end{abstract}

\section{Introduction}
Coherent charge transport  exploits the wave nature of the electrons in conductors to explore fundamental properties of charges in crystals. Innovative solid state devices and electronic circuits can be developed, in analogy to optical devices and quantum optics experiments, for applications in single-electron quantum optics, nanoelectronics and quantum technologies \cite{Forrester2023, Weinbub2022, Roussel2017, Rickhaus2015, Zhao2023, Adriaanse1991, Williams2011}. Quantum interference is one of the prototypical manifestations of the wave-like nature of electrons. Usually, interference effects are generated in devices where the electrons are spatially confined along specific directions, thus enabling the formation of optics-like interferometers having, for instance, Fabry-P\'erot \cite{Karalic2020, Liang2001, Kretinin2012} or Mach-Zehnder \cite{Ji2003, Huynh2012} geometry. Indeed, traditional electronic interferometers have been realized in low dimensional systems \cite{Ji2003, Bocquillon2013, Schuster1997, Fève2007, Karalic2020} or superconductors \cite{SQUID2004, vanDuzer2009}. 

The advent of graphene has opened new paths towards novel electron wave devices \cite{Liu2015, Wu2012, Young2009, Wilmart2014}. Compared to conventional two-dimensional electron gases, graphene offers easily tunable electrical properties and coherent transport accessible up to high temperatures \cite{Mayorov2011}. Several optics-like electronic elements have been currently realized \cite{Rickhaus2015,Handschin2017, Allen2017,Zhang2022, Liu2017}. The key element in graphene coherent devices is the external control of the local electrochemical potential, that allows the formation of energy barriers at the interface of regions of different doping (\emph{junction}). In the analogy with optics, the electrochemical potential plays the same role of the refractive index in ray optics when considering the semiclassical description \cite{Lee2015}, whereas the energy barriers act as semitransparent mirrors. Indeed, graphene potential barriers have electron transmission probability ($|t|^2$) that depends on the junction sharpness and the incident angle (Snell's law) \cite{Cheianov2006, Pereira2010}. In particular, graphene shows Klein tunneling, that is $|t|^2$ = 1 for orthogonal incidence \cite{Young2009, Pereira2010}. Based on these properties, Fabry-P\'erot-like resonant effects have been predicted \cite{Pereira2010} and observed in single or multiple gate graphene field-effect transistors. In particular, these phenomena manifest in the form of gate-induced oscillations in the total charge current flowing along the structure \cite{Allen2017, Wu2012, Drienovsky2014, Gehring2016, Varlet2014, Ahmad2019, Handschin2017, Rickhaus2013}. Indeed, when sweeping the gate voltage (i.e., the electrochemical potential), the Fermi wavelength of the charge carriers is tuned. When the Fermi wavelength matches the wavelengths of the allowed cavity modes, the charge carriers are spatially confined, unlocking resonant phenomena.\\
Resonant cavities can be realized by large potential steps, i.e., by junctions formed by regions with carriers of opposite polarity ($p-n$ junctions) or by high/low density of carriers of the same sign ($n^{--}-n$ or $p^{++}-p$ junctions). Generally, potential barriers in graphene are created at the metal/graphene interface due to the workfunction difference \cite{Wu2012} or by electrostatic gating \cite{Marconcini2019, Young2009, Huber2020, Dubey2013}. The former does not have tunable properties, the latter is effective mainly for $p-n$ junctions. Indeed, Fabry-P\'erot (FP) resonances have been predominantly observed in graphene transistors operating in unipolar electronic transport regime \cite{Allen2017, Wu2012, Gehring2016, Varlet2014, Ahmad2019, Handschin2017},  that is when the interference occurs only for one sign of the charge carriers. 
Moreover, gate-induced potential barriers require complex combinations of different nano-fabrication techniques to create ultra-narrow metal stripes and suffer from fringing electric field effects. The latter cause the enlargement of the effective barrier width and the smoothing of potential step sharpness \cite{Drienovsky2014, Dubey2013}, limiting the lateral resolution of the potential barriers and affecting the interference pattern, especially in multiple gate configuration. 

\begin{figure}[t]
\centering
 \includegraphics{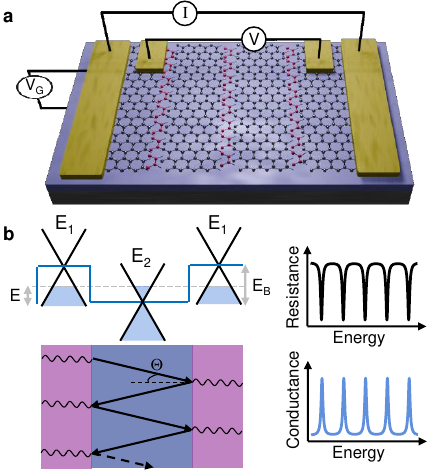}
\caption{\textbf{General description of the defective graphene FP interferometers.} \textbf{a} Sketch of interferometers, consisting in a periodic array of alternating quasi-ballistic (black atoms) and defective (pink atoms) graphene nano-stripes. The FP modes are probed by monitoring the graphene sheet resistance via four-probe measurements (see Methods). \textbf{b} Description of the resonant cavities formed by defective nano-stripes in graphene. Potential barriers of height $E_B$ ($E_B = |E_1 - E_2|$) are formed between defective (large hole-doping with chemical potential equal to $E_1$; pink areas) and quasi-ballistic (gate-tunable chemical potential $E_2$; violet area) nano-stripes. For example, when considering electron-doped quasi-ballistic nano-stripes, a resonant cavity is established: electrons are confined and reflected back and forward by the potential step at the interface with the defective nano-stripes. Only electrons with incident $\Theta \neq$ 0 can interfere due to the Klein tunneling. When sweeping the electrochemical potential $E$, dips (peaks) in the graphene resistance (conductance) are then expected at the energies of FP modes.}
\label{Fig1}
\end{figure}

Here, we demonstrate collective coherent effects induced in graphene by defect engineering. To this scope, we produced a simple charge carrier interferometer architecture based on the succession of pristine and defective graphene nano-stripes (Figure \ref{Fig1}a). Our approach makes use of artificially-created defects in the graphene lattice via electron-beam irradiation to generate charge carrier quantum interference with clear fingerprints in the device charge carrier transport. In other words, typically-undesired defects are here exploited to induce and regulate coherent phenomena in the charge carrier transport of graphene sheets, thus refuting the parallelism between defects and disorder. Moreover, our approach is based on \textit{direct writing}. It exploits the standard electron-beam lithography system, but it does not involve polymeric resist materials, thus reaching a cavity length scale that is not achievable via standard electrostatic-gating-based approaches.
Contrarily to the typical elastic scattering-induced localization effects obtained in disordered conductors \cite{Bluhm2019}, in our structure graphene defects simply introduce controlled potential barriers (Figure \ref{Fig1}b), while preserving the quasi-ballistic transport in the defect-free areas. Indeed, air exposed defective graphene can be strongly hole doped, thus operating as potential barrier. Consequently, our structure implements a FP charge carrier cavity by regularly alternating parallel defective, acting as mirrors, and quasi-ballistic nano-stripes with nanometric lateral resolution \cite{Melchioni2022}. When tuning the electrochemical potential, periodic oscillations in the sheet resistance of graphene are observed both for holes and electrons, demonstrating the $\emph{bipolar}$ operation regime. The FP modes are affected by possible disorder in the potential barrier heights, as confirmed by our simulations based on the Landauer-B\"uttiker formalism. Indeed, the simulations reproduce both the oscillatory behavior of the sheet resistance and the formation of multi-peak resonant modes in the presence of sizeable barrier inhomogeneity. Finally, temperature-dependent measurements show that the FP modes are robust up to 30 K, thus allowing us to extract the charge carrier coherence time, fundamental for the interferometer operation.

\section{Results and Discussion}

\subsection{Fabry-P\'erot charge resonances in defective graphene}
The FP interferometers are realized in monolayer graphene, which is mechanically exfoliated onto a doped-silicon substrate covered with 300 nm of thermally-grown silicon dioxide. Then, chromium/gold electrical contacts are fabricated by electron-beam lithography followed by thermal evaporation. Afterward, the graphene flakes are shaped in rectangles via reactive ion etching. Finally, the FP interferometer is fabricated by scanning a multi-line pattern with pitch $p$ and number of lines $N$ onto the graphene sheet with 20 keV electron-beam (see Experimental section for the details and Figure \ref{FigS1}, Supporting Information ). An effective linear density of $\sim$ 2$\times10^{6}$ cm$^{-1}$ defects of different nature is formed in each irradiated line, as estimated by micro-Raman spectroscopy  (Discussion 1, Supporting Information) \cite{Melchioni2022}. In general, these defects boost the physical adsorption and activate chemisorption of graphene \cite{Bhatt2022, Basta2023, Ye2017, Criado2015, Melchioni2022}, so that they locally alter the carrier density with respect to pristine graphene.  For example, the simple exposure to ambient atmosphere makes defective graphene largely hole-doped  \cite{Kumar2020}. By taking advantage of this simple doping mechanism, we form sharp potential barriers ($k_F d \ll 1$, where $k_F$ is the Fermi wavevector and $d$ is the barrier edge smearing \cite{Katsnelson2006, Liu2012}) at the interface between defective and pristine nano-stripes. By keeping the electron-exposed graphene for three days in ambient conditions, the irradiated lines gain large hole-doping of about 300 meV ($E_1$), whereas the unirradiated areas conserve the as-exfoliated doping of $\sim$ 200 meV ($E_2$), as extracted via micro-Raman spectroscopy ( Discussion 1, Supporting Information). The barrier height is quantified as $E_B = |E_1 - E_2| \sim 100$ meV. 
By this method, multiple parallel potential barriers are realized with a well-defined pitch of 50 nm and 75 nm \cite{Melchioni2022}. These barriers have Gaussian profile and full-width ($2\sigma$, where $\sigma$ quantifies also the barrier edge smearing $d$) of about 25 nm, as imposed by the electron-beam. Hence, the pristine nano-stripes have length $\ell = p - 2\sigma$ $\sim $ 25 nm (for $p$ = 50 nm) and $\sim $ 50 nm (for $p$ = 75 nm). 

\begin{figure*}[ht!]
\centering
\includegraphics{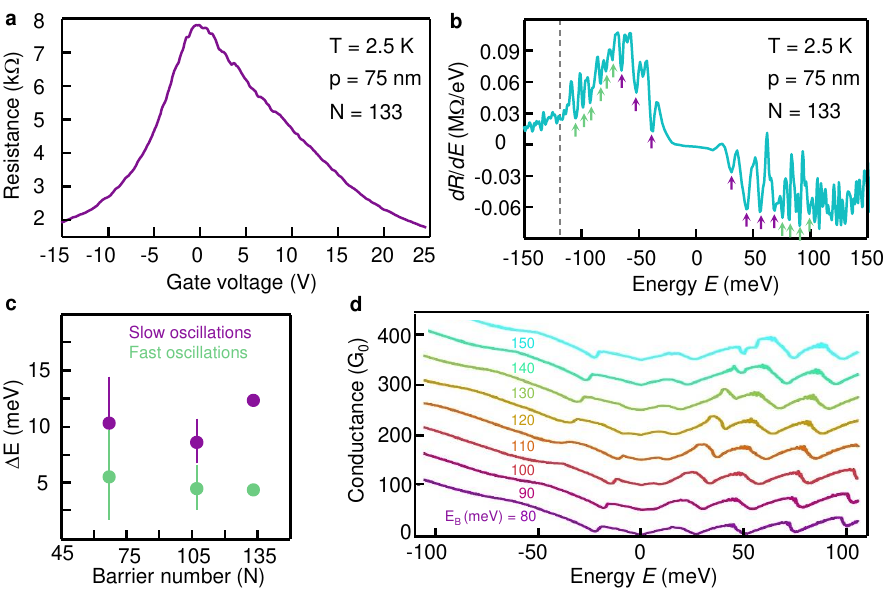}
\caption{\textbf{FP resonances in the graphene charge carrier transport.} \textbf{a} Graphene sheet resistance as a function of the gate voltage for multi-barriers with $p$ = 75 nm and $N$ = 133. The measurements are carried out in a four-probe configuration, injecting a source-drain current of 50 nA. The oscillations are observable in a large range of injected current, as shown in Supplementary Figure \ref {Figcurr}. \textbf{b} $dR/dE$ as a function of the energy $E$ calculated from the $R-V_G$ curve of panel \textbf{a}.  The violet arrows indicate the slow oscillations, while the green arrows the fast oscillations. The vertical dashed line stands for the energy where the resistance oscillations disappear in the hole-branch. This establishes the barriers height in the hole-transport, in agreement with Raman spectroscopy data. All measurements are carried out at 2.5 K. \textbf{c} Energy separation $\Delta E$ (estimated from the electron transport) for the slow (violet) and fast (green) oscillations as a function of the barrier number $N$. The error bar is quantified as maximum semi-dispersion, thus measuring the uniformity of the oscillations periodicity. The error bars for $N = 133$ are smaller than the symbols. Similar values are obtained in the hole-transport. \textbf{d} Conductance (in units of the conductance quantum $G_0$)  simulated via the Landauer-B\"uttiker formalism as a function of the energy for $N$ = 10 and barrier height $E_B$ ranging from 80 meV to 150 meV. The curves are vertically shifted for the sake of clarity.}\label{Fig2}
\end{figure*}

The charge transport properties of the FP interferometers are investigated in a four-probe configuration by injecting a constant current ($I$) through the device while recording the voltage ($V$) for different values of the back-gate voltage ($V_G$) (Figure \ref{Fig1}a). We note that this configuration is an accurate counterpart of the optical interferometry experiments, since a constant current keenly simulates the steady intensity of a laser beam.
Figure \ref{Fig2}a shows the bipolar and periodic oscillations in the resistance $R$ vs $V_G$ characteristics measured for a multi-barrier interferometer with $p$
= 75 nm and $N$ = 133. In particular, the device exhibits several dips superimposed onto an almost
standard $R - V_G$ curve (slightly electron/hole asymmetrical) for graphene.

When calculating the energy-to-resistance transfer function ($dR/dE$, where $E$ is the energy, i.e. the electrochemical potential), the oscillating behaviour becomes more explicit (Figure \ref{Fig2}b). In electron-dominated transport, $dR/dE$ shows slow oscillations with energy separation $\Delta E_{exp}$ of about 13 meV in the energy range close to the Dirac point, whereas it has fast oscillations with $\Delta E_{exp} \sim$ 5 meV for $E > $ 70 meV. This shorter $\Delta E_{exp}$ suggests a modification of the FP cavity possibly due to the detailed shape of the potential barrier with energy.
Analogous oscillations with similar periodicity occur in the hole transport branch, but, differently from the electron-side, they disappear at $E \sim $ -110 meV, i.e. when the electrochemical potential matches the barrier height.

The observed resistance oscillations represent the interference pattern of the electron (hole) waves propagating through the alternating $N$ defective and pristine nano-stripes. Indeed, two defective stripes and a pristine one build up a FP cavity (Figure \ref{Fig1}b). The 25 nm-long defective stripes act as partially reflecting mirrors capable to transmit phase-coherent carriers \cite{Childres2022}. Instead, the pristine zone supports a quasi-ballistic transport, thus it is the medium supporting the interfering carrier-waves (see Discussion 2, Supporting Information).
Experimentally, we observe that the periodicity of the resistance dips is independent of the number of barriers $N$ (Figure \ref{Fig2}c and Figure \ref{FigS3}, Supporting Information, for the full transport characterization), indicating that the $N$ FP cavities are independent but stringed together. Consequently, the measured interference pattern is averaged over $N$ FP interference patterns.

To get more insight into the resonant properties of the graphene FP interferometer, we numerically calculated the sheet conductance when varying the electrochemical potential by combining the envelope-function-based solutions of the Dirac-Weyl equation with the Landauer-B\"uttiker formalism \cite{Marconcini2022} (see Methods). These simulations assume identical equally-spaced Gaussian barriers with different height $E_B$ and fixed width 2$\sigma$. As shown in Figure \ref{Fig2}d, the conductance exhibits several peaks both in the hole- and electron-branch. Moreover, its dependence on energy is stronger for negative energies (hole-side), in full agreement with the experimental observations. In the Landauer-B\"uttiker approach, peaks in the conductance correspond to the maximum carrier transmission probability of the energy barriers. In our system, this is caused by the constructive interference among the charge waves. Thus, the values of $E$ at the conductance peaks (and thus resistance dips) correspond to the energies of the FP modes of the simulated interferometer. Generally, in monolayer graphene the separation energy  $\Delta E$ among the FP modes is defined by the cavity length $L_C$ as $\Delta E = {\pi \hslash v_F}/{L_C}$, where $\hslash$ is the reduced Planck constant and $v_F$ is the Fermi velocity. It is worth noting that in graphene Klein tunneling also affects the FP interference and its visibility: only the waves with non-normal incidence to the barriers ($\Theta \neq 0$, Figure \ref{Fig1}b), but with $\Theta$ such that both the transmission and reflection coefficients are reasonably high \cite{Allen2017}, participate in the resonances. For example, in sharp-edge barriers, $|t(\Theta)|^2 \sim$ 0.5 occurs only when $\Theta \sim 50^\circ$ for $|E| \sim E_B/2$ \cite{Katsnelson2006}. 
The presence of Klein tunneling implies that (i) a quasi-conventional $R$ versus $V_G$ characteristics is superimposed to the FP oscillations, as shown in Figure \ref{Fig2}a; (ii) the effective $L_C$ is longer than the distance between the potential barriers, i.e., along the orthogonal direction. 

The equally spaced FP modes of our simulations show a good qualitative and quantitative agreement with the experimental data. In particular, the separation energy of the simulated modes is $\sim$ 25 meV (the same order of the experimental $\Delta E_{exp} \sim$ 13 meV), corresponding to $L_{C, sim} \sim$ 85 nm and an effective incidence angle $\Theta_{sim}$ that is $\sim 55^\circ$ (see Discussion 3, Supporting Information ). By exploiting the same argument, $L_{C,exp} \sim$ 160 nm and $\Theta_{exp}$ $\sim 75^\circ$. It is worth noting that the simulations consider $N$ identical barriers and do not take into account the possible extension of the defects into the pristine stripes due to lattice re-arrangement, as it may occur in the experiments. These defects contribute to the quasi-ballistic transport as elastic (phase-preserving) scattering centers, making the cavity length longer compared to a defect-free stripe. Consequently, the longer the $L_C$, the smaller the $\Delta E$. Moreover, due to elastic scattering, the incidence angle at the barriers step can be also modified, and thus the barriers transmittance, influencing the interference pattern visibility.

\begin{figure*}[h]
\centering
\includegraphics {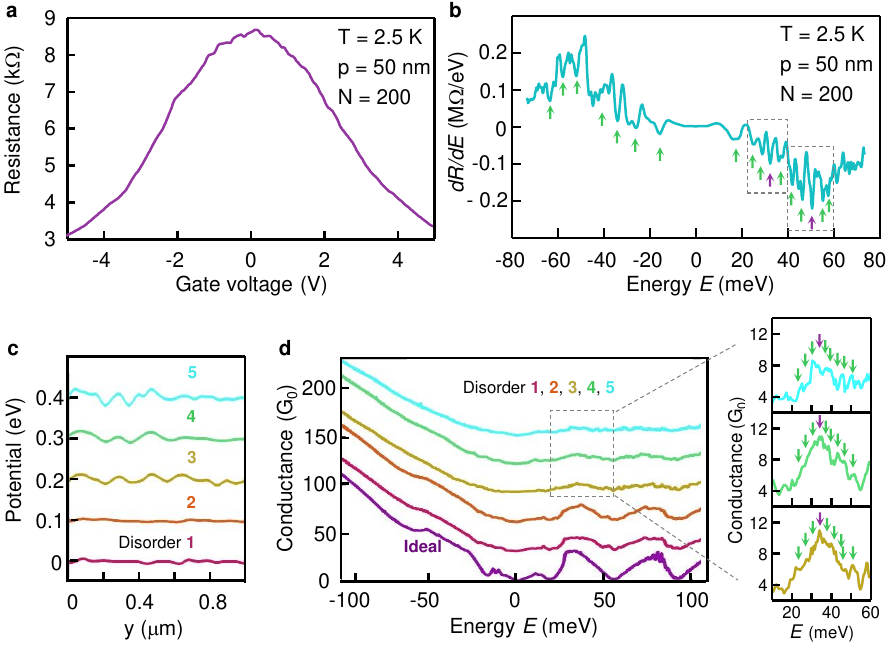} 
\caption{\textbf{Effects of the potential height disorder on the Fabry-P\'erot resonances}. \textbf{a} Sheet resistance as a function of the gate voltage for $p$ = 50 nm and $N$ = 200 measured at 2.5 K with 5 nA injected source-drain current. \textbf{b} $dR/dE$ as a function of the energy $E$ obtained from the $R-V_G$ curve of panel \textbf{a}. The green arrows indicate the fast resistance oscillations, whereas the violet ones and the dashed lines the slow oscillations. \textbf{c} Representative distribution of the potential barrier height along the direction parallel to the barriers for increasing disorder degree (disorder 1-5). The curves are the mean values of the curve reported in Figure \ref{FigS5}, Supporting Information. The curves are vertically shifted for the sake of clarity.
\textbf{d} Simulated sheet conductance (in units of the conductance quantum $G_0$) versus energy $E$ for $p$ = 50 nm and $E_B$ = 110 meV when introducing an increasing potential disorder (disorder 1-5) on the identical and homogeneous potential barriers (indicated as ideal). The disorder is randomly distributed onto the graphene sheet (Figure \ref{FigS5}, Supporting Information). The curves are vertically shifted for the sake of clarity. Inset: multi-peak resonances (green arrows) around the center of the main oscillation (violet arrows) caused by the potential disorder.}\label{Fig3}
\end{figure*}

\subsection{Effects of potential disorder and temperature}

When reducing the pitch down to 50 nm (thus $\ell \sim$ 25 nm), distinct oscillations in the sheet resistance are still clearly observable (see Figure \ref{Fig3}a), thus confirming the robustness of the proposed method for obtaining FP interferometers. In this case, the numerical simulations predict FP modes every $\sim$ 40 meV, corresponding to $L_{C, sim} \sim$ 55 nm and an effective $\Theta_{sim} \sim 60 ^\circ$ (similar to the cavity length/normal barrier distance ratio and $\Theta$ of $p$ = 75 nm). \\ 
The analysis of $dR/dE$ reveals that in the electron-branch the slow oscillations with $\Delta E_{exp} \sim$ 20 meV (violet arrows in Figure \ref{Fig3}b, corresponding to $L_{C, exp} \sim$ 105 nm and $\Theta_{exp}$ $\sim 70 ^\circ$) are split into fast oscillations with $\Delta E_{exp} \sim$ 4 meV (green arrows in Figure \ref{Fig3}b). Instead, in the hole-branch, only the fast oscillations can be clearly distinguished. This behaviour is reproduced also when probing a different number of barriers (see Figure \ref{FigThCyc}b, Supporting Information).

In order to understand the origin of these fast oscillations, we simulated the sheet conductance including randomly-distributed disorder in the potential barrier height (Figure \ref{FigS5}, Supporting Information). Indeed, experimentally potential inhomogeneities are unavoidable. This disorder can originate from the substrate electrostatics, carrier puddles, or local doping inhomogeneities within the defective stripes \cite{Martin2008}. The latter are more probable at short pitches because of the higher interaction among the irradiated stripes \cite{Melchioni2022}. The simulations consider different degrees of potential disorder (see Figure \ref{Fig3}c). As shown in Figure \ref{Fig3}d, this disorder strongly impacts the FP resonances. Differently from the ideal case (where the resonances are a set of well-defined spectral bands), the FP modes are less visible, when slowly increasing the potential fluctuations (disorder 1 - 2). For higher potential disorder (disorder 3 - 4), the main FP resonances are broader and less resolved. Moreover, they show superimposed fast oscillations with $\Delta E_{sim} \sim$ 5 meV (inset in Figure \ref{Fig3}d). Tentatively, these multi-peak resonances are explained by the fact that the disorder locally induces a slight variation of the potential barrier height $E_B$. Consequently, the FP modes energy is also modified due to their spectral dependence on $E_B$, as visible by comparing the curves at slightly different $E_B$ in Figure \ref{Fig2}d. Hence, the main multi-peak resonances arise from the spectral overlap of several FP modes having close resonance energies. For highly disordered barriers, instead, the main resonant modes and the fast oscillations almost completely disappear (disorder 5).

\begin{figure}[t]
\centering
\includegraphics{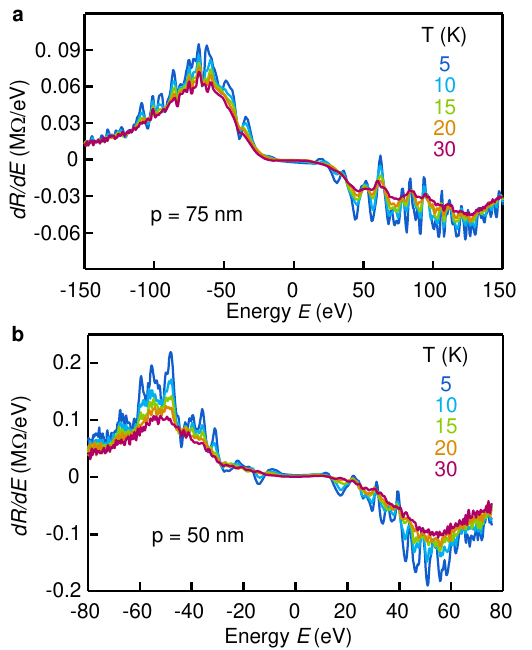}
\caption{\textbf{Temperature dependence of the resistance oscillations.} $dR/dE$ as a function of the energy for $p$ = 75 nm (\textbf{a}) and 50 nm (\textbf{b}) in the range of temperature $T$ between 5 K and 30 K.}\label{Fig4}
\end{figure}

Our graphene FP interferometers show resonant oscillations of resistance up to a temperature $T$ = 30 K (see Figure \ref{Fig4}). At all $T$, the resistance oscillations manifest at the same energies. By rising $T$, they are slowly suppressed, as expected by the thermal reduction of the charge carrier coherence length ($\ell_\phi$), until they disappear when $\ell_\phi$ becomes comparable to $L_C$. The coherence length is defined as $\ell_\phi = \sqrt{D \tau_\phi}$, with $D = (v_F \ell_{mfp})/2$ the diffusion constant, $v_F$ the Fermi velocity, $\ell_{mfp}$ the mean-free-path and $\tau_\phi$ the coherence time. In our interferometers, $D$ is about 0.02 m$^2$/s, since the lower bound of the mean-free-path is $\sim$ 30 nm (see Discussion 2, Supporting Information). Consequently, imposing $\ell_\phi$($T$ = 30 K) $\sim L_{C, exp}$, the maximum value of $\tau_\phi$ at 30 K can be quantified in $\sim$ 0.7 ps for $p$ = 50 nm and $\sim$ 2 ps for $p$ = 75 nm. It is worth noting that the longer $\tau_\phi$ for $p$ = 75 nm is reflected in the fact that the oscillations are not fully suppressed at 30 K, thus confirming our analysis. 

Finally, we stress that the FP interference pattern is reproducible over different thermal cycles, even the multi-peak resonance modes for $p$ = 50 nm (see Figure \ref{FigThCyc}, Supporting Information).
This feature fully confirms the quantum interference origin of the observed resistance oscillations and rules out other trivial explanations.

\section{Conclusions}
We presented an original approach to generate coherent quantum phenomena in graphene sheets by exploiting the wave nature of charge carriers. In particular, electron-irradiation-induced defects in the graphene lattice are exploited to mediate the coherent carrier transport by acting as partially transmitting mirrors, that confine the charge carriers within pristine quasi-ballistic nano-stripes (resonant structures). This capability strongly differs from the conventional role  of lattice defects as elastic scattering centres in diffusive conductors, where localization phenomena usually occur and dominate the charge transport. 
The regular alternation of defective and pristine nano-stripes creates $N$ independent FP cavities, whose interference fringes manifest as temperature-robust (up to 30 K) oscillations of the graphene sheet resistance. Moreover, these charge carrier interference patterns are present for both electrons and holes, revealing a bipolar behaviour, which is typically hardly observable in gate-induced barriers. As confirmed by numerical simulations, possible disorder in the barrier height generates multi-peak FP modes resulting in an increased number of resistance oscillations. Finally, the interference pattern is not correlated to $N$, indicating that thus the FP cavities are independent. 
Differently from the standard electrostatic approach based on gating, this methodology is extremely simple. On the one hand, it avoids the use of metal stripes that limit the minimum cavity length (typically $\ge$ 50 nm), offering the possibility to fully scale down and produce resonant structures also on a chip scale-level (if using high quality large-area graphene). On the other hand, thanks to the enhanced chemical reactivity, it offers an additional knob to control the coherent charge transport: when properly functionalized \cite{Basta2023}, the defective graphene allows to define \textit{ad hoc} the properties of the potential barriers. Consequently, electron-irradiated graphene represents an original route to develop innovative coherent electronics devices for the manipulation of carrier waves and the manifestation of quantum charge-optics effects.

\section {Experimental section} \label{Methods}
\subsection*{Graphene interferometers fabrication}
Monolayer graphene flakes are mechanically exfoliated (natural graphite  from NGS Trade and Consulting GmbH and blue tape from Nitto Italia srl) onto doped-silicon substrates covered with 300 nm of thermally-grown silicon dioxide. The substrate chips are initially cleaned by oxygen plasma at 100 W for 5 minutes to promote the graphene adhesion and remove organic residues from the substrate surface to maximize the defect patterning lateral resolution \cite{Basta2022}.

First, electrical contacts to the graphene sheet are fabricated by combining electron-beam lithography with thermal evaporation of a 5 nm/50 nm-thick film of chromium/gold. The graphene flakes are subsequently dry-etched via reactive ion etching method with a polymer mask to obtain the rectangular shape (10 $\mu$m-wide and different length from 2 $\mu$m to 10 $\mu$m). Finally, the chips are deeply cleaned by resist remover (All resist, AR 600-71) to eliminate the residues from the graphene surface.

The defective nano-stripes are produced by scanning a multi-line pattern on the graphene surface with an electron-beam at 20 keV. The pitch between the defective lines is $p$ = 50 nm and $p$ = 75 nm, whereas the irradiation step-size along each line is 7.8 nm. The number of defective stripes are $N$ = 40, 200 for $p$ = 50 nm and $N$ = 67, 107, 133 for $p$ = 75 nm. The electron beam is produced by a commercial lithography system (from Raith GmbH) and it has a beam spot of $\sim$ 25 nm. For all patterns, the electron-beam current is about 100 pA, delivering a linear dose of $\sim$ 31 mC/cm and resulting in a dwell-time of $\sim$ 250 $\mu$s.\\
Prior to the electrical measurements, the irradiated samples are exposed to ambient atmosphere for three days in order to hole-dope the defective stripes and establish the potential barriers. 

The crystal lattice of the irradiated-graphene is studied in ambient by micro-Raman spectroscopy (Renishaw). The Raman spectrum is acquired by scanning the graphene surface with a laser at 532 nm with a 100$\times$ objective (NA = 0.85), that corresponds to a lateral resolution $<$ 1 $\mu$m. The map step-size is 500 nm.  The laser power is set at 118 $\mu$W to exclude any possible laser heating of the lattice. More details about the defective multi-stripe topography and crystallographic structure are reported in Ref. \cite{Melchioni2022}.\\ 

\subsection*{Charge transport measurements}
The interferometric properties of the periodically-irradiated graphene is measured in a dry cryostat (OptiStat Dry by Oxford Instruments), having a base temperature of 2.5 K. The measurements as a function of temperature are carried out by employing a temperature controller (Mercury iTC by Oxford Instruments), which allows the sweep from 2.5 K up to 300 K. The temperature stability of the system is better than 50 mK.

The electronic transport of the carrier-interferometers is measured in a four-probe configuration. An AC source-to-drain current at 17 Hz is injected into the graphene and the voltage drop across two probe contacts is measured via lock-in technique (Stanford SR830). The DC gate voltage (sweep step of 100 mV) is applied by a source-meter (Keithley 2614B) in back-gate configuration.\\

\subsection*{Charge transport simulations}
In order to keep the computational time within reasonable limits but without loss of generality, we simulated a 1~$\mu$m-wide graphene sheet containing 10 tunnel barriers, that are orthogonal to the transport direction. The barriers have a Gaussian profile along the transport direction with peak amplitude $E_B$ in the range 80-150 meV (10 meV step) and full-width 2$\sigma \sim$ 25 nm. We assume the distance between the barrier peaks to be either 50 nm or 75 nm.

The simulations are performed by combining the solution of the Dirac-Weyl envelope-function equation with the Landauer-B\"uttiker formalism \cite{Marconcini2019,Marconcini2021,Marconcini2022}. \\
In the low-energy range, the graphene electron-wave function can be written in terms of four envelope functions, $F_A^{\vec K}$, $F_B^{\vec K}$, $F_A^{\vec K'}$, $F_B^{\vec K'}$, each one corresponding to one of the two triangular sublattices $A$ and $B$ and of the two unequivalent Dirac points $\vec K$ and $\vec K'$. These four envelope functions have to satisfy the Dirac-Weyl equation, with Dirichlet boundary conditions enforced at the sheet edges. \\
The transport problem is turned into a set of much simpler ones using a recursive scattering matrix approach. The graphene sheet is partitioned into a series of slices, that are as wide as the graphene sheet (transverse direction, $y$), but with the length along the transport direction ($x$) sufficiently short to make the variation of the potential energy $U$ along the $x$-direction negligible within each slice. As a consequence, within each slice the longitudinal wave vector $\kappa_x$ is conserved and each envelope function $F (x,y)$ can be factored into a part $\Phi (y)$, that depends only on the transverse coordinate $y$, and into a plane wave $e^{i \kappa_x x}$ in the transport direction. Therefore, within each slice the Dirac-Weyl equation becomes the following differential eigenproblem:
\begin{equation}\label{system}
\left\{
\begin{aligned}
&-\big(\sigma_x f(y)+\sigma_z\partial_y\big)\, 
\vec{\varphi}^{\vec K}(y)=\kappa_x \vec{\varphi}^{\vec K}(y)\\
&-\big(\sigma_x f(y)-\sigma_z\partial_y\big)\, 
\vec{\varphi}^{\vec K'}(y)=\kappa_x \vec{\varphi}^{\vec K'}(y)\\
&\vec\varphi^{\vec K}(0) =\vec\varphi^{\vec K'}(0)\\
&\vec\varphi^{\vec K}(W) =
e^{i 2 K W} \vec\varphi^{\vec K'}(W)\,.\\
\end{aligned}
\right.
\end{equation}
where $\vec\varphi^{\vec K}(y)=[\Phi^{\vec K}_A(y),\Phi^{\vec K}_B(y)]^T$, $\vec\varphi^{\vec K'}(y)=i\,[\Phi^{\vec K'}_A(y),\Phi^{\vec K'}_B(y)]^T$, $\sigma_x$ and $\sigma_z$ are the Pauli matrices, $\partial_y=d/d\,y$, $f(y)=[U(y)-E]/[\hbar v_F]$ (with $v_F$ the graphene Fermi velocity, $\hbar$ the reduced Planck constant, $E$ the electron energy), $K=|\vec K|$, and $W$ is the effective width of the sheet. In order to optimize the calculations and avoid the fermion doubling problem, which arises from a standard direct space discretization of the Dirac equation, the problem is transformed into this equivalent one, as follows
\begin{equation}\label{system1}
\left\{
\begin{aligned}
& \!-\!\!\big(\sigma_z\partial_y\!+\!\sigma_z i K \!+\!
\sigma_x f(W\!-\!|W\!-\!y|)\big)\vec\rho(y)\!=\!\kappa_x \vec\rho(y)\\
& \vec\rho(2W)=\vec\rho(0)\,,
\end{aligned}
\right.
\end{equation}
where the function $\rho(y)$ is defined in the range $[0,2W]$ as
\begin{equation}\label{rho}
\vec\rho(y)\!=\!
\left\{
\begin{array}{l l}
\!\! \vec\varphi^{\vec K}(y) e^{-iKy} &
\hbox{ for $0 \le y \le W$}\\
\!\! \vec\varphi^{\vec K'}(2W\!-\!y) e^{i K (2W-y)} &
\hbox{ for $W \le y \le 2W$}.
\end{array}
\right.
\end{equation}
The new differential problem (Equation \ref{system1}) has periodic boundary conditions on $\vec\rho(y)$ and thus it can be numerically solved (with a proper frequency cut) in the reciprocal space, obtaining a basis of solutions of the Dirac-Weyl equation. These solutions represent the transport modes in the slice.

Once the transmission modes of two adjacent slices are computed, the scattering matrix connecting these modes is obtained by mode-matching technique. For each mode, we write the total wave function (in terms of the unknown reflection and transmission coefficients), that results from the injection of that single mode into the pair of slices. This wave function has to satisfy the continuity condition at the interface between the two slices. By projecting the resulting set of continuity equations onto a basis of sine functions, a linear system in the reflection and transmission coefficients is thus obtained. Solving this system, the scattering matrix of the two coupled slices is calculated. Recursively composing the scattering matrices that connect all the slices in the sheet, the overall scattering matrix, and, in particular, the transmission matrix $t$ between the modes in the input and output leads, is finally computed. From this matrix, the conductance $\mathcal{G}$ of the graphene can be then estimated by exploiting the Landauer-B\"uttiker formula as
\begin{equation}\label{landauer}
\mathcal{G}=\frac{2\,e^2}{h}\sum_{n,m} |t_{nm}|^2\,,
\end{equation}
where $e$ is the elementary charge and $h$ is the Plank constant. The sum is performed over all the modes that propagate in the input and in the output leads, which are specified by the indices $m$ and $n$, respectively.

Finally, the disorder in the potential height is reproduced \cite{Marconcini2021} by adding to the potential profile a sum of two-dimensional Gaussian functions, i.e., $\tilde V_p\,e^{-\frac{\delta^2}{2 \tilde \sigma^2}}$, where $\delta$ is the distance from the Gaussian center and $\tilde \sigma$ = 35.7 nm. These Gaussian functions are randomly distributed over the graphene sheet with a concentration equal to $10^{10}$~cm$^{-2}$ and differ in peak amplitude ($\tilde V_p$). $\tilde V_p$ is randomly chosen, with a uniform distribution, between $-15.5$ and $+15.5$ meV.\\

\section*{Acknowledgements}
The authors thank S. Cassandra for fruitful discussions. The authors wish to acknowledge C. Coletti from the Istituto Italiano di Tecnologia for the access to the micro-Raman facility.

F. B. and F. P. acknowledge partial financial support under the National Recovery and Resilience Plan (NRRP), Mission 4, Component 2, Investment 1.1, Call PRIN 2022 by the Italian Ministry of University and Research (MUR), funded by the European Union – NextGenerationEU – EQUATE Project, "Defect engineered graphene for electro-thermal quantum technology" - Grant Assignment Decree No. 2022Z7RHRS.

F. B. acknowledges partial financial support by the
MUR - Italian Minister of University and Research under the “Research projects of relevant national interest - PRIN 2020” - Project No. 2020JLZ52N, title “Light-matter interactions and the collective behavior of quantum 2D materials (q-LIMA)”.

S.R. acknowledges partial financial support under the National Recovery and Resilience Plan (NRRP), Mission 4, Component 2, Investment 1.1, Call PRIN 2022 by the Italian Ministry of University and Research (MUR), funded by the European Union – NextGenerationEU – TRUST project "Trampolines as ultra-Sensitive thermomechanical bolometers" - Grant Assignment Decree No. 2022M5RSK5.

P. M. and M. M. acknowledge partial financial support by the Italian Ministry of University and Research (MUR) in the framework of the FoReLab and CrossLab project (Department of Excellence) and by the European Union, Next-GenerationEU, National Recovery and Resilience Plan (NRRP), Mission 4 Component 2 Investment N. 1.4, CUP N. I53C22000690001, through the National Centre for HPC, Big Data and Quantum Computing (“Spoke 6: Multiscale Modelling $\&$ Engineering Applications” and “Spoke 10: Quantum Computing”).

This material is based in part upon work supported by the U.S. Department of Energy, Office
of Science, National Quantum Information Science Research Centers, Superconducting
Quantum Materials and Systems Center (SQMS) under contract number DE-AC02-07CH11359.

\section*{Conflict of interest}
The authors declare no conflict of interest.

\section*{Data availability}
The data that support the findings of this study are available in this article and its Supporting Information. Any other relevant data are available from the corresponding authors upon reasonable request.

\section*{Contributions}
N.M. fabricated the devices and carried out the micro-Raman characterization. N.M. and F.P. performed charge carrier transport experiments with inputs from F.B.. P.M. and M.M. conducted numerical simulations. N.M., F.P., S.R., A.T., and F.B. analysed and interpreted the data. F.B. conceived the idea of the project, coordinated research efforts and wrote the paper. All authors discussed the data and participated in preparing the manuscript.

\newpage
\begin{center}
  \section*{SUPPORTING INFORMATION}
\date{} %%%date is omitted
\maketitle   
\end{center}

\section*{Figure S1: Description of the graphene periodical irradiation}

\begin{figure}[h]
\begin{center} 
\includegraphics {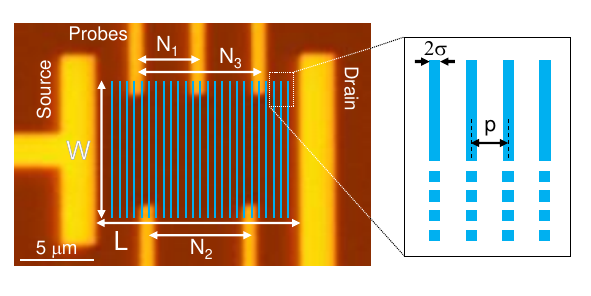}
\caption{\textbf{Description of the graphene periodical irradiation.} The graphene is shaped as a rectangle $W$-wide and $L$-long. Electrical contacts are defined for four-probe measurements: source and drain for current injection and 5 probes for voltage drop measurements at different lengths. The defective stripes are regularly spaced with $p$-pitch on the graphene sheet (cyan lines). Each stripe is 2$\sigma$-wide. The probe contacts investigate the charge carrier transport through $N_i$-potential barriers ($i$ =1, 2, 3), where $N_1 = 67$, $N_2 = 107$ and $N_3 = 133$ for $p$ = 75 nm and $N_1 = 40$ and $N_2 = 200$ for $p$ = 50 nm.} \label{FigS1}
\end{center}
\end{figure}

\newpage
\section*{Discussion 1: Doping and defect probing via micro-Raman spectroscopy}
\begin{figure}[h]
\begin{center} 
\includegraphics{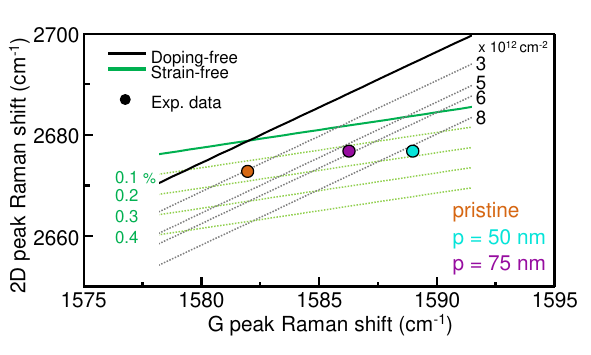} 
\caption{\textbf{Doping probing via micro-Raman spectroscopy.} Correlation plot that allows to disentangle the strain and the doping contribution from the 2D and G peak Raman shift. The dots are the experimental data. The solid black and green lines describe the functional dependency of the 2D peak Raman shift on the G peak Raman shift in doping- and strain-free conditions, respectively. The black dotted lines are the linear dependence of the 2D peak Raman shift versus the G peak Raman shift at different doping  when varying the tensile strain. The dotted green lines are the linear dependence of the 2D peak Raman shift versus the G peak Raman shift at different tensile strain values when changing the charge doping.} \label{FigRaman}
\end{center}
\end{figure}

A quantitative analysis of the charge doping and the density of defects induced in the lattice via electron-irradiation can be extrapolated from micro-Raman spectroscopy data. \\
\vspace{0.5cm}
\textbf{Doping probing}\\
In the low defect density regime, the correlation plot method from Ref. \cite{Lee2012} is valid \cite{Eckmann2013, Lucchese2010} and thus, the charge doping can be estimated. As reported in Figure \ref{FigRaman}, the two-dimensional (2D) carrier density $n_{2D}$ is quantified as $\sim 3 \times 10^{12}$ cm$^{-2}$ for pristine graphene (from the as-exfoliated values \cite{Basta2022, Melchioni2023}), $n_{2D}$ $\sim 6 \times 10^{12}$ cm$^{-2}$ for $p$ = 75 nm, and $n_{2D}$ $\sim 8 \times 10^{12}$ cm$^{-2}$ for $p$ = 50 nm.

\begin{figure}
\begin{center} 
\includegraphics{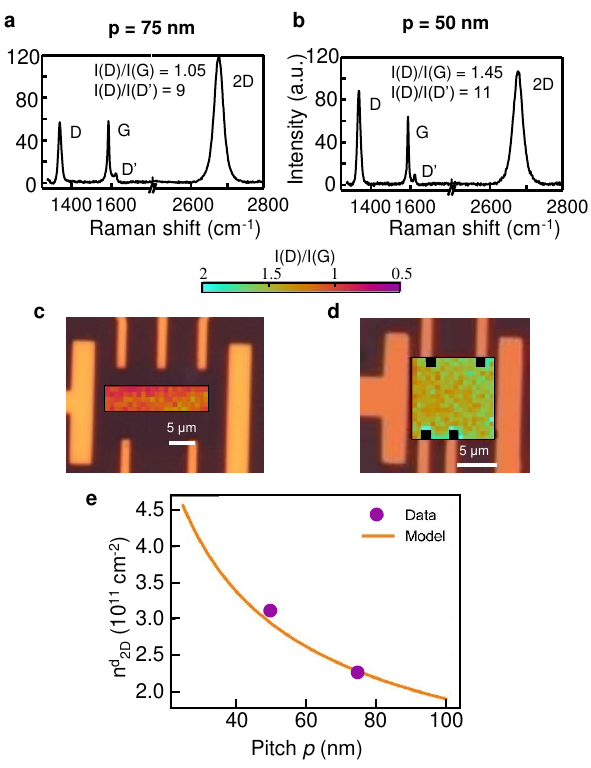} 
\caption{\textbf{Defects probing via micro-Raman spectroscopy.} Averaged Raman spectrum for devices with $p$ = 75 nm (\textbf{a}) and $p$ = 50 nm (\textbf{b}). Two-dimensional spatial distribution of $I(D)/I(G)$ for $p$ = 75 nm (\textbf{c}) and $p$ = 50 nm (\textbf{d}) overlapped on the corresponding device optical images. \textbf{e} 2D defect density $n^d_{2D}$ as a function of the pitch. The violet dots are the experimental data and the orange line is the predicted $p$-dependence of $n^d_{2D}$ from the phenomenological model introduced in Ref. \cite{Melchioni2022}.} \label{FigS2}
\end{center}
\end{figure}

\textbf{Defects probing}\\
As shown in Figure\ref{FigS2}a-b, the Raman spectrum exhibits the $D$ and $D'$ bands. These bands are the unambiguous signature of defects in the crystal and their analysis gives insight into both the nature \cite{Eckmann2012} and the density of the defects \cite{Bruna2014}.

The nature of the defects in the crystal can be identified from the ratio between the $D$ and the $D'$ peak intensities ($I$). The $I(D)$/$I(D') \sim$ 11 (for $p$ = 75 nm) and 9 (for $p$ = 50 nm) indicates mostly sp$^3$-like defects, as expected for defective graphene exposed to air atmosphere for several days \cite{Melchioni2023, Eckmann2012}. \\
Instead, the two-dimensional (2D) sheet density of defects ($n^d_{2D}$) is computed correlating the intensities of the D and G peaks [$I(D)$ and $I(G)$, respectively]. According to the formula \cite{Bruna2014}
\begin{equation} \label{eqn2d}
n^d_{2D} \mathrm{[cm^{-2}]} = 7.3\times10^9 E_L^4\mathrm{[eV^4]} \frac{I(D)}{I(G)}
\end{equation}
where $E_L$ is the pump laser energy (in our case, 2.33 eV), we obtain a 2D defect density of $n^d_{2D} \sim 2.3\times 10^{11}\ \mathrm{cm^{-2}}$ for $p = 75$ nm [uniformly distributed $I(D)/I(G) \sim$ 1.05,  Figure \ref{FigS2}c] and $n^d_{2D} \sim 3.1\times 10^{11}\ \mathrm{cm^{-2}}$ for $p = 50$ nm [uniformly distributed $I(D)/I(G) \sim$ 1.45, Figure \ref{FigS2}d].
As discussed in the main text, the samples consist of periodic repetitions of defective and pristine nano-stripes. The laser spot size is $\sim$ 1 $\mu$m, so that it cannot spatially resolve the defective and pristine areas. This means that the acquired Raman spectrum is an average spectrum containing information from multiple defective and pristine nano-stripes. To give an estimation of the effective density of defects along each nano-stripe, we consider that, for a given pitch, the number of lines per unit length is $\rho_{lines} = 1/p$. Therefore, $n^d_{2D}$ can be defined in terms of one-dimensional (1D) density $n^d_{1D}$ as follows
\begin{equation}\label{eqn1d}
n^d_{2D} = n^d_{1D}\rho_{lines} = n^d_{1D}/p
\end{equation} 

Consequently, combining Eq. \ref{eqn2d} with Eq. \ref{eqn1d}, we estimate an effective 1D density of defects $n^d_{1D} = 1.69\times 10^{6}\ \mathrm{cm^{-1}}$ and $n^d_{1D} = 1.56\times 10^{6}\ \mathrm{cm^{-1}}$ for $p = 75$ nm and 50 nm, respectively. 

Interestingly, the values of $n_{1D}$ are not constant [although all the lines are exposed with the same beam parameters (see Methods in the main text)], indicating that the $n_{2D}$ dependence on the pitch is different from a simple $n_{2D}\propto p^{-1}$. This is explained by the presence of a cross-talk among the defective nano-stripes. This cross-talk arises from proximity effects occurring during the irradiation process, as explained more in details in Ref. \cite{Melchioni2022}, where the phenomenological model describing the dependence of the $I(D)/I(G)$ values on the pitch $p$ is introduced, that is 
\begin{equation} \label{model}
\frac{I(D)}{I(G)} = \frac{241}{p \mathrm{[nm]}}\cdot \left(\frac{p\mathrm{[nm]}}{1500}\right)^{0.37}
\end{equation}

By combining Eq. \ref{eqn2d} with Eq. \ref{model}, the expected $n_{2D}$ versus $p$ can be retrieved. As shown in Figure \ref{FigS2}e, the model well describes the estimated values.

\newpage
\section*{Figure S4: FP resonances at different source-drain currents}
\begin{figure}[h]
\begin{center} 
\includegraphics{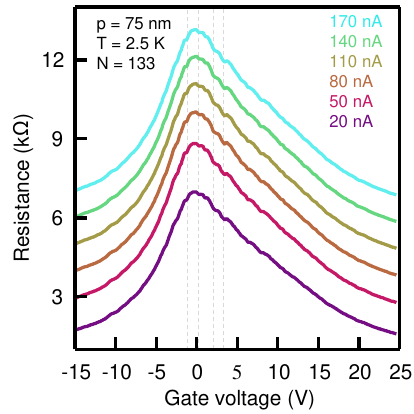} 
\caption{\textbf{Measured sheet resistance at different injected source-drain currents.} Sheet resistance at various injected currents from 20 nA (bottom) to 170 nA (top) measured at 2.5 K for $p$ = 75 nm and $N$ = 133. The curves are vertically shifted for the sake of clarity. The FP interference does not change by varying the current injection. The vertical dashed grey lines mark some of the resistance oscillations.} \label{Figcurr}
\end{center}
\end{figure}

\newpage
\section*{Discussion 2: Phase-coherent quasi-ballistic transport through periodically irradiated graphene sheet}
For $p$ = 75 nm (50 nm), the pristine stripes have a length of 50 nm (25 nm). Here, a coherent quasi-ballistic transport can be safely assumed. Generally, the mean-free path ($\ell_{mfp}$) can be estimated from the mobility, which is extracted by fitting the total measured sheet resistance \cite{Kim2009}. When focusing on the more symmetric hole-dominated transport (Figure \ref{Fig2}.a and \ref{Fig3}.a in the main text), the mean-free path is quantified as high as $\sim$ 30 nm. This is a lower bound value, as the measured resistance arises from the series of the defective and pristine graphene resistances. Consequently, quasi-ballistic trajectories are preserved, ensuring phase-matching among the carriers that travel back and forward within the pristine stripes after the reflection at the potential step built at the interface with the defective stripes. \newline
In the defective stripes, quantum interference effects are excluded. These stripes only take part in creating the $``$mirrors$"$, while quasi-ballistically transmitting phase-coherent carriers into the next nearest FP cavity. Based on the Raman spectroscopy results, the disorder in defective stripes is in the low density regime \cite{Eckmann2013, Lucchese2010}. In this regime, the carrier phase coherence length is almost unaffected, as demonstrated in similar electron-irradiated graphene \cite{Childres2022}. Instead, the mean-free-path becomes shorter compared to the pristine lattice \cite{Childres2022}, confirming our underestimation of $\ell_{mfp}$ in the pristine stripes. %However, the short length of the defective stripes (25 nm) is comparable with the $\ell_{mfp}$, thus some carrier trajectories is expected to remain (locally) ballistic. 

\newpage
\section*{Figure S5: Carrier transport through different number of potential barriers}

\begin{figure}[h]
\begin{center} 
\includegraphics{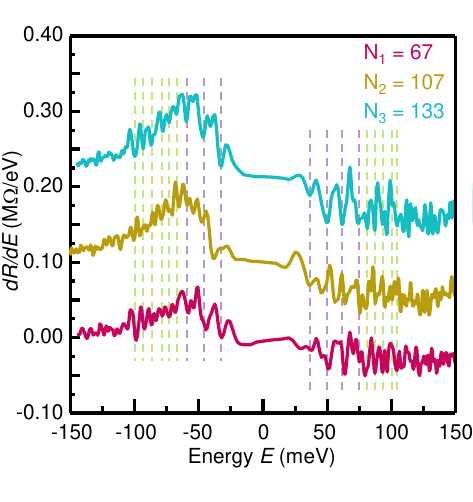} 
\caption{\textbf{Dependence of the resistance oscillations on the potential barrier number.} $dR/dE$ versus energy measured at 2.5 K for $p$ = 75 nm and $N_1$ = 67 (violet curve), $N_2$ = 107 (dark yellow curve), and $N_3$ = 133 (cyan curve). The vertical dashed lines indicates the slow (violet) and fast (green) oscillations of the resistance. The curves are vertically shifted for the sake of clarity. The number of barriers does not change the energy of the resonant modes, thus the device behaves as $N$ independent FP interferometers.} \label{FigS3}
\end{center}
\end{figure}

\newpage
\section*{Discussion 3: Charge carrier effective incident angle at the potential step}

\begin{figure}[h]
\begin{center}
\includegraphics{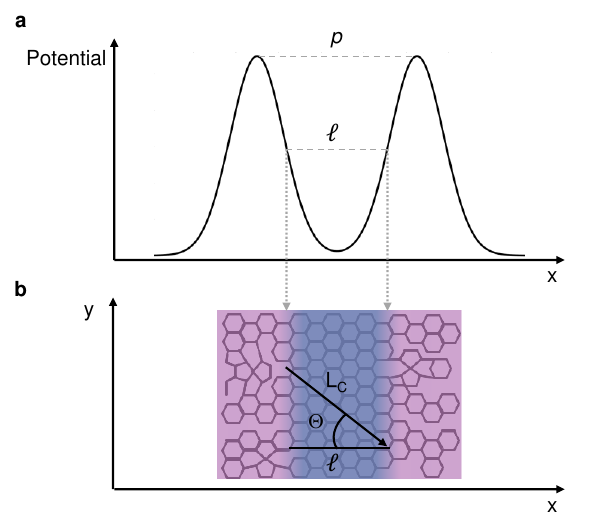}
\caption{\textbf{Effective incident angle $\Theta_{sim/exp}$ estimation.} \textbf{a} Spatial distribution of the barriers potential along the x-axis parallel to the source-drain current direction. \textbf{b} Scheme for the trigonometrical calculation of the charge carrier incident angle $\Theta$ at the potential step between the pristine (violet area) and the defective (pink areas) nano-stripes on the graphene plane.} \label{FigTheta}
\end{center}
\end{figure}

As explained in the main text, the presence of Klein tunneling imposes that the FP resonances are visible only when charge carriers have an incident angle $\Theta$ at the potential step such that the transmission and reflection coefficients are maximized. In order to give the estimation of an effective $\Theta$ both in the experiments and in the simulations, we assume a simplified picture (Figure \ref{FigTheta}) and we apply some trigonometrical calculations. Considering the ideal case where the pristine zone has length equal to $\ell$ = $p$ - 2$\sigma$ and by using the cavity length $L_C$ extracted from $\Delta E_{sim/exp}$, the incident angle $\Theta_{sim/exp}$ is quantified as
\begin{equation}
    \Theta_{sim/exp} = \arccos(\frac{\ell}{L_{C, sim/exp}})
\end{equation}

\newpage
\section*{Figure S7: Thermal cycle effects on the FP resonances}

\begin{figure}[h]
\begin{center}
\includegraphics{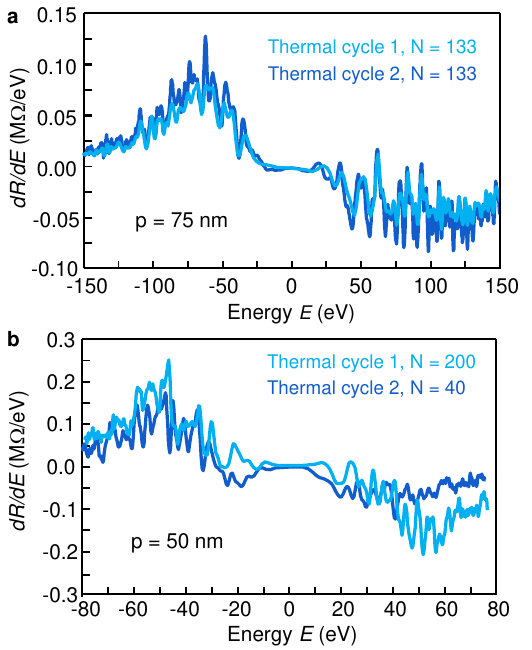}
\caption{\textbf{$dR/dE$ for different thermal cycles.} $dR/dE$ versus energy $E$ for $p$ = 75 nm (\textbf{a}) and 50 nm (\textbf{b}) measured at 2.5 K after two different thermal cycles. It is important noticing that for $p$ = 50 nm the resistance is measured through two different barrier numbers: $N$ = 200 and $N$ = 40 via two different probe contact pairs, demonstrating also for this pitch that the FP cavities are independent.}\label{FigThCyc}
\end{center}
\end{figure}

\newpage
\section*{Figure S8: Randomly-distributed disorder in the potential barrier height}

\begin{figure}[h]
\begin{center}
\includegraphics{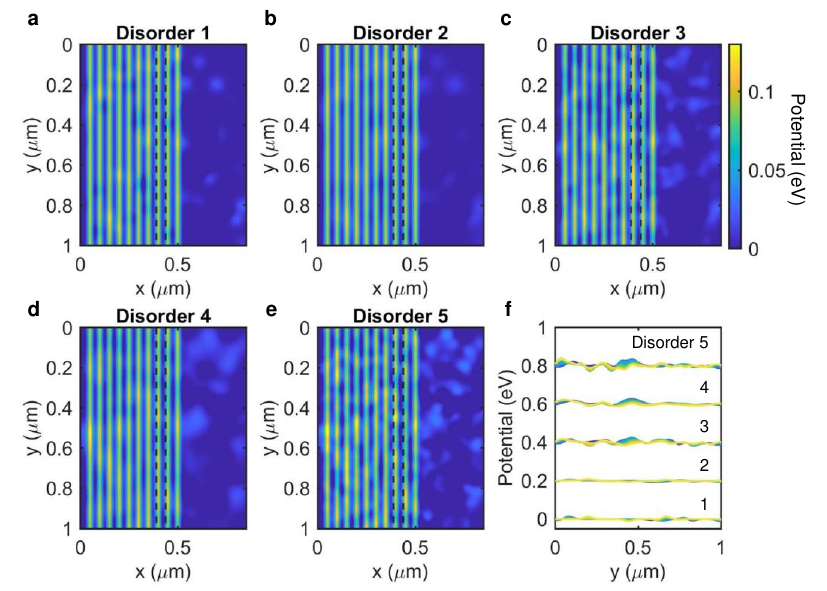}
\caption{\textbf{Barrier potentials in the presence of disorder.} \textbf{a - e} Two-dimensional map distribution of the potential barriers with different disorder degree for $p$ = 50 nm. \textbf{f} Line profile of the barrier potential along the y-axis. The shown values are those in the range indicated by the vertical dashed black lines in panels a - e. The curves are vertically shifted for the sake of clarity.}\label{FigS5}
\end{center}
\end{figure}

\end{document}